# Probing Bidirectional Plasmon-Plasmon Coupling-Induced Hot Charge Carriers in Dualplasmonic Au/CuS Nanocrystals

*Patrick Bessel[§], André Niebur[§], Daniel Kranz, Jannika Lauth\*, Dirk Dorfs\**

§ Both authors contributed equally to this work.

Patrick Bessel, André Niebur, Daniel Kranz, Jannika Lauth, Dirk Dorfs
Institute of Physical Chemistry and Electrochemistry, Leibniz Universität Hannover, D-30167 - Hannover, Germany
E-mail: dirk.dorfs@pci.uni-hannover.de
        jannika.lauth@uni-tuebingen.de

Patrick Bessel, Daniel Kranz, Jannika Lauth, Dirk Dorfs
Laboratory of Nano and Quantum Engineering, Leibniz Universität Hannover, D-30167 Hannover, Germany

André Niebur, Jannika Lauth, Dirk Dorfs
Cluster of Excellence PhoenixD (Photonics, Optics and Engineering – Innovation Across Disciplines), D-30167 Hannover, Germany

Jannika Lauth
Institute of Physical and Theoretical Chemistry, University of Tübingen, Auf der Morgenstelle 18, D-72076 Tübingen, Germany

Keywords: LSPR, localized surface plasmon resonance, transient absorption spectroscopy, hot charge carrier relaxation, ultrafast dynamics

Heterostructured Au/CuS nanocrystals (NCs) exhibit localized surface plasmon resonance (LSPR) centered at two different wavelengths (551 nm and 1051 nm) with a slight broadening compared to respective homostructured Au and CuS NC spectra. By applying ultrafast transient absorption spectroscopy (TAS) we show that a resonant excitation at the respective LSPR maxima of the heterostructured Au/CuS NCs leads to the characteristic hot charge carrier relaxation associated with **both** LSPRs in both cases. A comparison of the dualplasmonic heterostructure behavior with a colloidal mixture of homostructured Au and CuS NCs shows that the coupled dualplasmonic interaction is only active in the heterostructured Au/CuS NCs. By investigating the charge carrier dynamics of the process, we find that the observed interaction is charge carrier based as it is faster than phononic or thermal processes ($< 100$ fs). The relaxation of the generated hot charge carriers is faster for heterostructured nanocrystals, also indicating, that the interaction occurs as an energy transfer or charge carrier transfer between both materials. Our results strengthen the understanding of multiplasmonic interactions in heterostructured Au/CuS NCs and will significantly advance applications where these interactions are essential, such as catalytic reactions.

## 1. Introduction

Localized surface plasmon resonance (LSPR) is a phenomenon well described for nanocrystals of different – typically metallic – materials. LSPR is the resonant excitation of a collective oscillation of free charge carrier density with the incident light.[1,2] The wavelength of an LSPR mainly depends on the density and effective masses of the free charge carriers as well as on the permittivity of the surrounding medium.[1] By influencing these quantities the spectral position can be controlled, which is possible via doping[3], size and shape control[4,5], shell growth[6] or redox chemistry[7]. The combination of two plasmonic materials is gaining popularity as alternative to tune the LSPR, but is also to study LSPR interactions.[8–10] Combining two plasmonic materials with a similar LSPR frequency in heterostructured NCs typically leads to a combined LSPR band of both materials centered between the LSPRs of the respective single material NCs.[11] If the resonance frequencies of the two materials are sufficiently far apart, two separate LSPRs will also be apparent in the combined heterostructure.[8,9,12] For example, this is the case in gold and copperselenide ($Cu_{2-x}Se$, berzelianite) core-shell NCs, in which the resulting gold LSPR is centered in the visible wavelength range (578 nm) and the $Cu_{2-x}Se$ LSPR is centered at near infrared wavelengths (996 nm).[8] Such a system is of particular interest since the LSPR in $Cu_{2-x}Se$ is based on the collective oscillation of defect electrons. Possible interactions between the two LSPRs can be investigated by conventional steady state spectroscopy (UV-Vis-NIR absorption), which yields information about the spectral position and broadening of the corresponding absorption band. However, with a lifetime of a few femtoseconds, LSPR is a short-lived process.[5,13,14] The study of its temporal evolution or the temporal evolution of plasmon-induced hot charge carriers is therefore essential to draw conclusions about the underlying processes in a dual plasmonic system.

Generally, the LSPR dephases within a few femtoseconds leading to a nonequilibrium distribution of the excited charge carriers.[13,15] Charge carrier scattering (thermalization) within 100 fs results in a Fermi distribution referred to as hot electrons or hot holes, no longer considered as plasmons.[16] The hot charge carriers then can scatter with phonons on a timescale of up to 1 ps to 10 ps followed by a final relaxation through heat dissipation (phonon-phonon scattering) within a few hundred picoseconds.[14,15] A transfer of energy rich, hot charge carriers to molecules on the surface or domains of other materials during their relaxation is possible if the energy levels of the excited charge carriers and the receiving material are aligned.[15,17] This transfer gets more efficient with a smaller radius of the nanocrystals, due to scattering of the charge carriers with the surface, known as Landau damping.[18] In our investigated Au/CuS NC system, a cross excitation of gold and coppersulfide LSPRs is indicated by a transient change in optical properties operational in both directions. This also leads to small, but unspecific changes in the steady state spectroscopy.[8,9]

Heterostructured NCs represent a combination of different materials in a single NC and exhibit an excellent opportunity to combine properties of the different material combinations or to investigate interactions.[19] Dualplasmonic NCs with a spectral separation of both LSPRs like the Au/CuS system studied here have been synthesized successfully previously.[9,20] However, the interactions between both domains and their transient optical changes have not been investigated. Research on cross-interactions in dualplasmonic systems are still at an early stage. For example, Shan *et al.* have synthesized Au@$Cu_{2-x}$Se nanocrescents and performed TAS with a non-resonant excitation at 380 nm and 800 nm to investigate the underlying charge carrier interactions between Au and $Cu_{2-x}$Se and the changed optical properties of the heterostructured NCs.[8]

Here, we investigate synthesized dualplasmonic Au/CuS NCs with transient absorption spectroscopy to gain insights into cross-interactions between the electron LSPR at UV-Vis wavelengths and a hole LSPR in the near-infrared (NIR) wavelengths. The respective LSPRs are excited resonantly at 551 nm (2.25 eV) and 1051 nm (1.18 eV) and are probed with a low intensity white-light continuum covering 350 nm to 1400 nm (3.54 - 0.89 eV). The transient response is followed up to 300 ps after excitation and compared to the reference systems of homostructured Au NCs, CuS NCs, and a mixture of Au NCs and CuS NCs. Strikingly we find a typical hot charge carrier relaxation at LSPR wavelengths of **both** materials regardless which LSPR is excited resonantly and conclude, that energy and/or charge carrier transfer between both materials occurs, since the observed effect is faster than phononic or thermal processes. This is the first time that such type of interaction is observed and will help to deepen the understanding of all types of dualplasmonic materials.

## 2. Results and Discussion

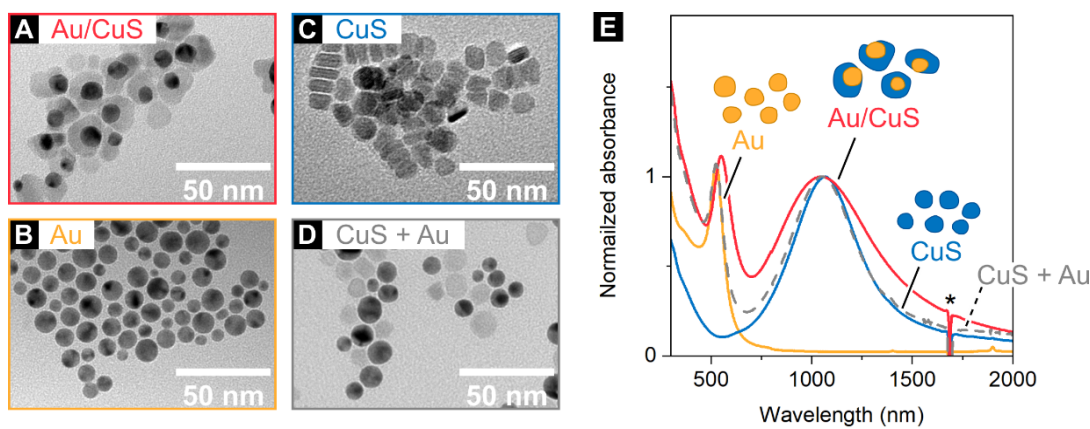

Figure 1: (A – D) TEM bright field images of (A) pure Au NCs, (B) pure CuS NCs, (C) heterostructured Au/CuS NCs and (D) a mixture of Au and CuS NCs. (E) Normalized absorbance spectra of the respective NCs dispersed in toluene. The spectra are normalized to the maxima of the CuS LSPR, except for the pure Au NCs which are normalized to have the same extinction as the Au LSPR of the NC mixture. *The sharp signal at 1685 nm originates from the baseline correction of the first harmonic of the aromatic C-H vibration absorption of toluene.[21]

**Figure 1**A-D shows TEM images of the synthesized pure Au NCs and CuS NCs respectively and heterostructured Au/CuS NCs. Heterostructured Au/CuS NCs exhibit a diameter of 19.6 nm ± 2.6 nm with a quasi-spherical Au core with a diameter of 9.4 nm ± 2.6 nm (*N*=250). With 83.4 % the majority of NCs is heterostructured (*N*=350). The remaining NCs are either pure CuS NCs (16.3 %) or pure Au NCs (0.3 %). These results are expected, as the heterogeneous nucleation is energetically preferred to the homogeneous nucleation in general.[22] The synthesized heterostructured Au/CuS NCs are Janus particles since the Au cores are not fully enclosed by the CuS shell. In the mixture (Figure 1D) consisting of pure Au NCs and pure CuS NCs used for our spectroscopic investigations, the Au NCs exhibit a diameter of 9.4 nm ± 2.2 nm and the CuS NCs have a diameter of 11.2 nm ± 1.4 nm. The mixture is composed of 50 % CuS NCs and 50 % Au NCs, heterostructured NCs are not found

by TEM analysis (*N*=350). All NCs are clearly separated from each other (see **Figure S1** in the Supporting Information for the size distributions). The molar Cu:Au ratio is 1.6 in the heterostructured NCs and 1 in the NC mixture determined with atomic absorption spectroscopy. X-ray diffraction patterns are shown in **Figure S2** for the pure NCs (Au yellow and CuS blue), the heterostructured NCs (red) and the NC mixture (gray) and match the reflections of elementary gold (PDF Card # 01-071-3755) and covellite (PDF Card # 00-006-0464) with minor impurities. The reflections are broadened as expected for nanomaterials. UV-Vis-NIR spectra of the NC dispersions are shown in Figure 1E: Both, the pure Au (yellow) and CuS NCs (blue) exhibit a single LSPR band with a maximum at 525 nm and 1067 nm, respectively. The heterostructured Au/CuS NCs (red) exhibit two maxima of the LSPR bands centered at at 551 nm and 1051 nm, respectively. The Au LSPR in the at UV-Vis wavelengths is bathochromically shifted by 26 nm, probably stemming from the higher permittivity at the Au LSPR frequency of the surrounding inorganic shell, which was calculated from the refractive index *n* for Au and CuS *via* an approximated Sellmeier equation ($\varepsilon_m(\text{toluene}) = 2.38$[23], $\varepsilon_{m,\,551\,\text{nm}}(\text{covellite}) \approx n_{551\,nm}^2 \approx 2.54^2 \approx 6.4$ [24]).[1] In contrast, the CuS LSPR is hypsochromically shifted by 16 nm, which is likewise explained by a change in permittivity of the surrounding medium, as the permittivity of gold ($\varepsilon_{m,\,1051\,nm}(\text{gold}) \approx n_{1051\,nm}^2 \approx 0.1^2 \approx 0.01$[25]) is lower than the permittivity of toluene.[1] Similar trends are found in other recent work.[8,9]

To answer the question if an interaction in Au and CuS NCs is only possible with a shared interface (like in the heterostructured Au/CuS NCs), a mixture of Au and CuS NCs is prepared for comparison with a similar ratio of extinction at the LSPR maxima in steady-state absorption as the heterostructured Au/CuS NCs. The LSPR maxima of the NC mixture are located at 525 nm and 1067 nm respectively and indicate that the pure NCs are not interacting with each other in the mixture. We derive the concentrations of Au and Cu from atomic absorption spectroscopy, the molar extinction coefficients are then calculated using the Beer-Lambert law. The extinction coefficient of the heterostructured Au/CuS NCs are lower than the values of the pure Au and CuS NCs or the mixture, matching previous reported results.[8] All dispersions are stable over several months when stored under inert conditions and show no changes in the UV-Vis-NIR spectra within this time period.

In order to gain information about the interactions between both LSPRs we use transient absorption spectroscopy (TAS). Heterostructured Au/CuS NCs are photoexcited resonantly at the maximum of the Au or the CuS plasmon at 551 nm (2.25 eV) or 1051 nm (1.18 eV), resp. The monoplasmonic systems and the mixture are used as references, on which the same TAS measurements are performed. We do not find any transient response for a non-resonant excitation of pure Au NCs at 1051 nm and pure CuS NCs at 551 nm as is shown in **Figure S3**. For the mixture of pure Au and CuS NCs we only observe the transient response of the hot charge carriers stemming from the resonantly excited LSPR, as shown in **Figure S4**. This observation is expected, since the absorption maximum of the pure Au NCs coincides with the absorption minimum of the pure CuS NCs and vice versa. When pure Au NCs are resonantly excited at 2.25 eV (551 nm), the typical transient signals are probed and shown in **Figure 2**A. A decaying bleach signal at 2.36 eV (525 nm) is hypsochromically shifted by 3 meV (0.6 nm) with respect to the maximum of the steady-state absorption and exhibits two adjacent induced absorption signals at 2.09 eV (593 nm) and 2.67 eV (464 nm). 10 ps after excitation, the bleach signal has decayed by 80 %. As expected, no absorbance change (nor a steady-state absorption) is probed in the NIR region. We observe a similar behavior for the resonant excitation of the CuS NCs at 1.18 eV (1051 nm) shown Figure 2B. Like in Au NCs, a resonant photoexcitation of the CuS NCs causes a decaying bleach signal at 1.09 eV (1137 nm), with an adjacent induced absorption signal in the visible range at 2.34 eV (529 nm). We expect a second induced absorption signal under 1 eV, since the adjacent positive signals are caused by the broadening of the absorption band.[26]

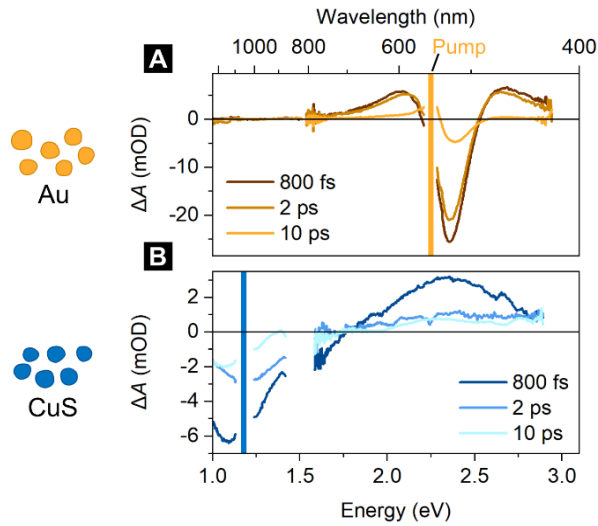

Figure 2: TAS spectral slices at 800 fs, 2 ps, and 10 ps after resonant excitation of (A) pure Au NCs (yellow) and (B) pure CuS NCs (blue) dispersed in toluene. Vertical coloured lines are wavelengths of the pump pulses at 2.25 eV (551 nm) and 1.18 eV (1051 nm), resp.

By following the evolution of the transient signals over time, insights are gained into the processes described above (dephasing, thermalization, charge carrier-phonon scattering, and phonon-phonon scattering). However, the dephasing and charge carrier thermalization step is not observed directly here because of temporal resolution restraints. First, the generated laser pulses have a pulse duration of 100 fs and the IRF for the measurements is limited to ~150 fs. Second, the pulses propagate through 2 mm of toluene containing the NCs. Therefore, the pulses experience chromatic dispersion of toluene, which leads to an additional loss of temporal resolution, especially when probed in the visible range and pumped in the NIR (1.18 eV).[27] The cooling processes of plasmon-induced hot electrons and hot holes however exhibit time constants between several 100 fs and several 100 ps and can therefore be followed nicely by the measurement. For analysis of the pure Au NCs TAS measurement, we use the measured signal, which is the excited absorption $A^*$ subtracted by the ground state absorption $A_0$. Our fitting procedure is presented for the simplest case: the resonant excitation of pure Au NCs and is applied to all other cases subsequently. The spectral slices are fitted as the difference of two Gaussian curves representing $A^*$ and $A_0$, as shown in **Figure 3**A.

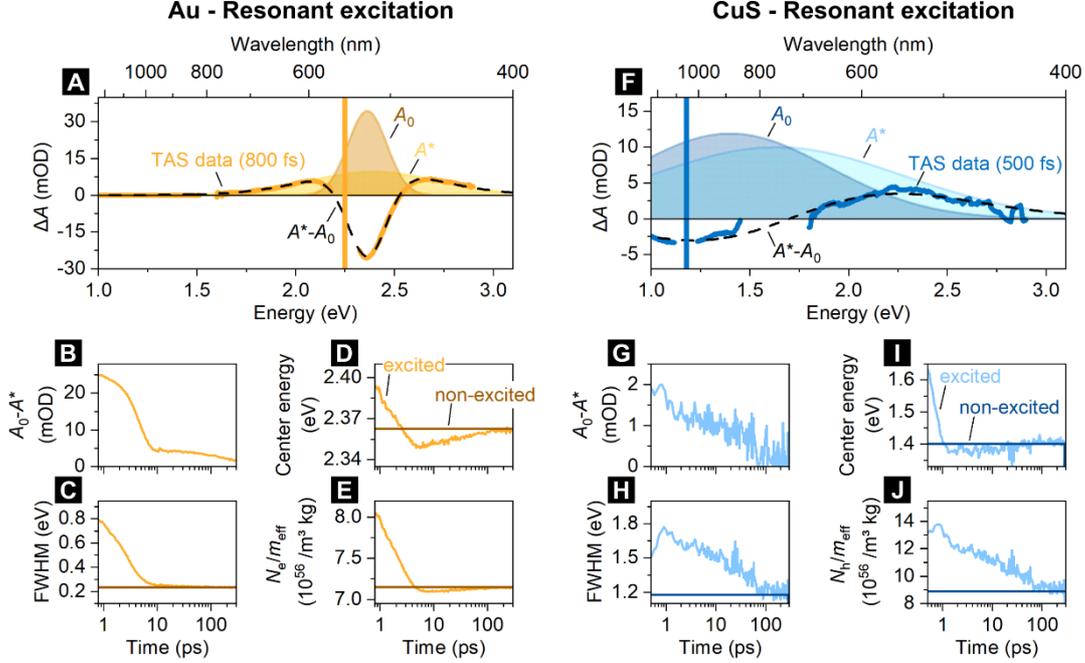

Figure 3: (A) TAS of resonantly excited Au NCs after 800 fs. The dotted line indicates the fitted data as a difference of the excited and a non-excited contribution ($A^*$, yellow and $A_0$, brown, respectively). (B-D) Fitted amplitude, center energy, and FWHM plotted *vs.* time after the excitation. (E) Charge carrier density over effective charge carrier mass *vs.* the time derived from the fitted parameters. (F-J) Same values as in panels A-D for resonantly excited CuS NCs.

The data is well described by this approach, even though interband transitions occur at higher energies, which are visible above 3.0 eV (e.g. transition 5d to 6sp band with 3.1 eV, in general >2.4 eV)[28] in the steady-state absorption spectra in Figure 1A. We assume that these interband transitions are not excited by the pump pulse or that possible changes are significantly faster than the instrument response function (IRF) of our measurement (150 fs to 350 fs). We find no transient signals indicative for these transitions. The measured $\Delta A$ spectrum is exclusively induced by the presence of hot charge carriers and the associated decrease of charge carriers in the ground state levels. The vanishing of the absorption change over time is caused by the relaxation of the hot charge carriers to the ground state. This is taken into account by keeping $A_0$ fixed over time and recalculating $A^*$ for each time step, as shown in **Figure S5**. As $\Delta A$ approaches zero over time, the excited state $A^*$ also converges into the ground state $A_0$. This fitting procedure provides the time evolution of the amplitude (*Amp*), the energetic position and the broadening after excitation, which all approach the respective ground state parameters, see Figure 3B-D. The progression of the fit parameters, with the non-excited absorbance displayed in brown and the evolving excited state absorbance in yellow, respectively. Further conclusions are drawn from the fit parameters by applying the Drude model, which is commonly used to analyse LSPRs bands in steady-state spectroscopy.[1]

$$\omega_{\mathrm{sp}} = \sqrt{\frac{\omega_{\mathrm{p}}^2}{\varepsilon_\infty + 2\varepsilon_{\mathrm{m}}} - \gamma^2} \qquad\qquad (1)$$

where $\omega_{\mathrm{sp}}$ is the LSPR frequency, $\omega_{\mathrm{p}}$ is the bulk plasma frequency, $\varepsilon_\infty$ is the relative permittivity at high frequencies ($\varepsilon_\infty(\mathrm{Au}) = 1.53$[29], $\varepsilon_\infty(\mathrm{CuS}) = 8.4$[30]), $\varepsilon_{\mathrm{m}}$ is the relative permittivity of the surrounding medium, i.e. toluene ($\varepsilon_{\mathrm{m}} = 2.38$[23]), and $\gamma$ is the damping factor determining the linewidth (FWHM) of the LSPR.[1] The frequency of the plasmon $\omega_{\mathrm{p}}$ is equal to

$$\omega_{\mathrm{p}} = \sqrt{\frac{N\,q_{\mathrm{e}}^2}{\varepsilon_0 m_{\mathrm{eff}}}} \qquad\qquad (2)$$

where $N$ is the density of the charge carriers, $q_{\mathrm{e}}$ is the elementary charge, $\varepsilon_0$ is the vacuum permittivity and $m_{\mathrm{eff}}$ is the effective mass of the respective charge carriers.[1] By substituting this expression into Equation (1) and transforming it, an expression for the ratio of the density of the charge carriers and their effective mass $N/m_{\mathrm{eff}}$ is obtained:

$$\frac{N}{m_{\mathrm{eff}}} = \frac{(\omega_{sp}^2 + \gamma^2)(\varepsilon_\infty + 2\varepsilon_m)\varepsilon_0}{q_e^2} \qquad\qquad (3)$$

The expression for $N$ (Equation (3) multiplied by $m_{eff}$) is commonly applied to LSPR bands by assuming a constant $m_{eff}$.[13] For TAS measurements, however, a constant effective mass of the charge carriers cannot be assumed, as the electronic structure is drastically altered by the pump pulse. An expression for $N/m_{eff}$ takes into account that a change of both the charge carrier density and the effective mass determine the respective LSPR.[1,31] In all derived data (amplitude, energetic position, broadening and $N/m_{eff}$), shown in Figure 3B-E, from the TA signals of the pure Au and CuS NCs, we find two mono-exponential decay contributions before the ground state is reached again. This corresponds to a fast electron-phonon scattering process and a slower phonon-phonon scattering process, summarized in Table (1). Except for the amplitude, all parameters are well described by bi-exponential decay dynamics. The fitted kinetics are shown in **Figure S6**.

Table 1: Time constants of the amplitude, center energy, FWHM, and $N/m_{eff}$ fitted by bi-exponential decays, derived from resonantly excited Au NCs and CuS NCs respectively. *The fast decay process doesn't follow an exponential decay.

| Analysed parameter | Pure Au NCs | | Pure CuS NCs | |
|---|---|---|---|---|
| | $\tau_{fast}$ (ps) | $\tau_{slow}$ (ps) | $\tau_{fast}$ (ps) | $\tau_{slow}$ (ps) |
| Amplitude | * | $215 \pm 25$ | $0.8 \pm 0.2$ | $63 \pm 11$ |
| Center energy | $1.43 \pm 0.03$ | $30 \pm 2$ | $0.27 \pm 0.02$ | $26 \pm 7$ |
| FWHM | $2.26 \pm 0.02$ | $32 \pm 3$ | $4.7 \pm 2.3$ | $53 \pm 8$ |
| $N/m_{\text{eff}}$ | $1.60 \pm 0.02$ | $20 \pm 2$ | $1.4 \pm 0.3$ | $49 \pm 6$ |

For resonantly excited Au NCs we assign the faster relaxation process of $N/m_{\text{eff}}$ with a time constant of $(1.60 \pm 0.02)$ ps to carrier-phonon and the slower process with $(20 \pm 2)$ ps to phonon-phonon scattering. These values are in good agreement with previously reported time constants measured with TAS (2.5 ps, and > 50 ps).[14] Measurements and data evaluation of resonantly excited CuS NCs are shown in **Figure 3F-J**. The fitting procedure of CuS NCs is less accurate compared to the Au NCs because the TAS measurement spans the full available spectral range and one of the two induced absorption signals cannot be mapped due to the probing range limit at 1600 nm. Therefore, relatively strict boundary conditions were set for the fitting of the first time step, as summarized in Table S1. We find that $A^*$ shifts *via* two processes associated with fast hole-phonon scattering and slow phonon-phonon scattering with time constants of $(270 \pm 20)$ fs and $(26 \pm 7)$ ps, respectively. The time constant of carrier-phonon scattering is consistent with literature observations (< 1 ps[32]) and is generally faster than in noble metal NCs (1 ps to 4 ps[33]). The remaining parameters of $A^*$ (Amp, FWHM, $N/m_{\text{eff}}$) do not follow a clear two-phase relaxation, which is attributed to the rather low fit quality of the spectral slices. For heterostructured Au/CuS NCs we follow the same procedure by resonantly photoexciting the structures either at the Au or the CuS LSPR band maximum. When exciting resonantly to the energetically higher Au LSPR band maximum (2.25 eV), we observe transient changes both at visible and NIR wavelengths see **Figure 4**A. In fact, the transient features of pure Au NCs and CuS NCs are found in the $\Delta A$ spectrum of the heterostructure Au/CuS NCs when excited at 2.25 eV. This observation implies the presence of hot electrons and holes associated with the high-energy Au LSPR and low-energy CuS LSPR, respectively, and an energy transfer process between the Au and CuS domain. However, interestingly, the **same observation** is made when the heterostructured Au/CuS NCs are excited resonantly to the CuS LSPR, as shown in Figure 4B.

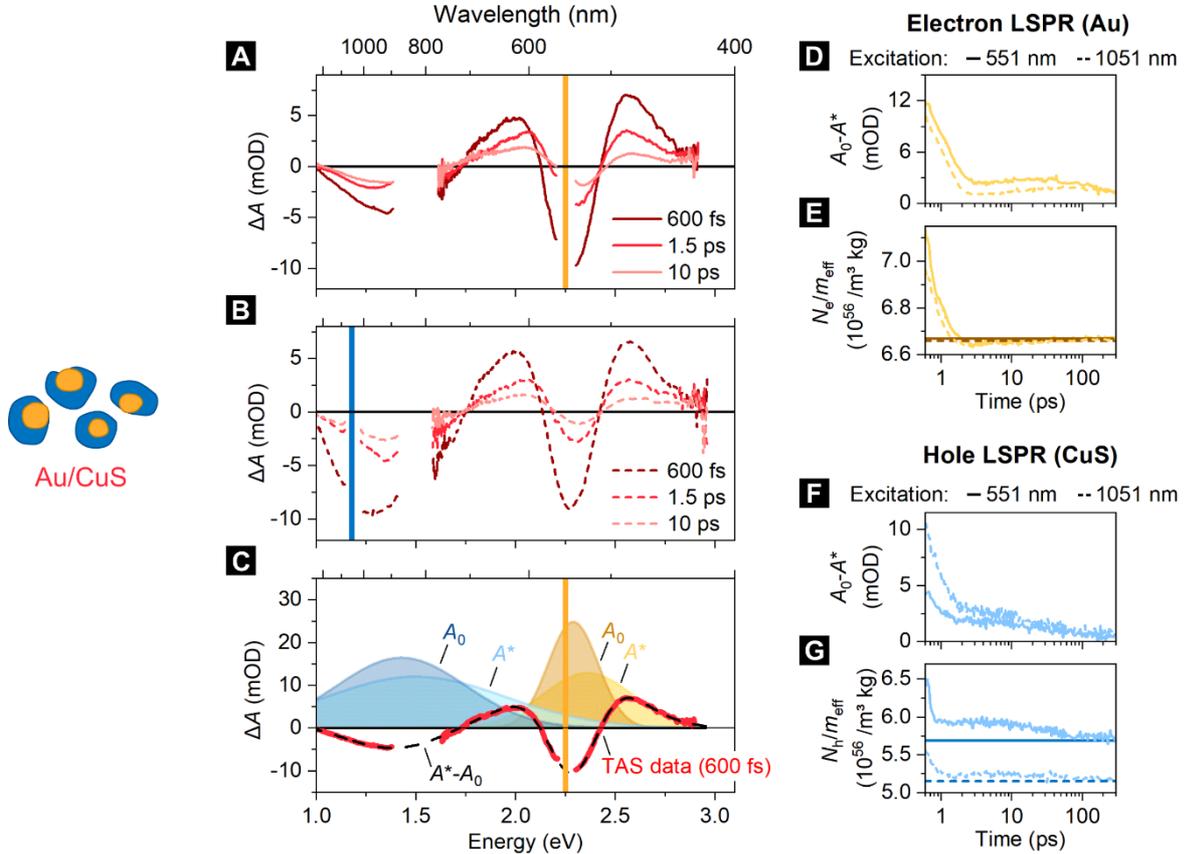

Figure 4: (A, B) TAS of heterostructured Au/CuS NCs at 600 fs, 1.5 ps, and 10 ps after excitation with 551 nm and 1051 nm pulses, respectively. (C) Transient absorption of Au/CuS NCs at 600 fs after excitation with a 551 nm pulse. (D-F) Fitted amplitude, center energy, and FWHM versus time after excitation corresponding to the electron plasmon. The solid lines represent photoexcitation at 551 nm and the dashed lines 1051 nm. (G) Charge carrier density over effective charge carrier mass of the electron plasmon versus time calculated with fitted parameters. (H-K) The same as in panels D-G but for the CuS plasmon.

The energy of the pump pulse (1.18 eV) is not sufficient to directly excite the LSPR of pure Au NCs, as shown in **Figure S3**A. Nevertheless, we probe the characteristic transient response of hot electrons in the visible wavelength range. Since we probe transient responses in both energetic regions, the data is fitted using four Gaussians, which represent the excited and ground state absorption. We find almost identical decay dynamics regardless which LSPR band in heterostructured Au/CuS NCs is excited resonantly, as is shown for the amplitude and $N/m_{\text{eff}}$ for both excitation energies in **Figure 4**F-G, and for the spectral position and FWHM in **Figure S**6. Note that the fitting procedure of the hot electrons (Au domain) leads to decay dynamics that are almost independent of the excitation energy of the pump pulse. Meanwhile, all fitting parameters of the hot holes (CuS domain) have dynamics that are independent of the excitation energy but have constant offsets. Again, these offsets are likely linked to the difficult fitting procedure of the transient signal of the hot holes, being overlayed by the response of the hot electrons in the visible range. Since the decay dynamics of both, the hot electrons and the hot holes are independent of whether the electron plasmon or the hole plasmon was excited, we assume that the same thermalized system is present 600 fs after excitation at 1.18 eV or 2.25 eV. This implies, that the underlying transfer processes between the Au and CuS domain are faster than the respective IRFs between 150 fs and 300 fs and are therefore assumed to originate from charge carrier interactions, since both carrier-phonon and phonon-phonon scattering processes are orders of magnitude slower.[15] In contrast the time constant of the decaying hot electrons in pure Au NCs is slower ((1.60 ± 0.02) ps) than in the heterostructured NCs ((0.36 ± 0.01) ps) again indicating a fast charge carrier based interaction between the Au and the CuS compartment, which allows the carriers generated in the gold to relax partially via the CuS compartment (which has a faster carrier-phonon relaxation dynamics). The time constants of the fitted amplitude and N/m$_{\text{eff}}$ for both excitation energies are summarized in Table 2.

Table 2: Time constants of the amplitude, and $N/m_{\text{eff}}$ fitted by bi-exponential decays, derived from heterostructured Au/CuS NCs that are resonantly excited to the Au LSPR (2.25 eV) and the CuS LSPR (1.18 eV) respectively.

| | | Hot Electrons | | Hot Holes | |
|---|---|---|---|---|---|
| Excitation (eV) | Parameter | $\tau_{\text{fast}}$ (ps) | $\tau_{\text{slow}}$ (ps) | $\tau_{\text{fast}}$ (ps) | $\tau_{\text{slow}}$ (ps) |
| 1.18 | Amplitude | $0.74 \pm 0.02$ | $4.0 \pm 0.4$ | $0.45 \pm 0.02$ | $32 \pm 3$ |
| 2.25 | Amplitude | $0.66 \pm 0.01$ | $520 \pm 40$ | $0.36 \pm 0.02$ | $39 \pm 4$ |
| 1.18 | $N/m_{\text{eff}}$ | $0.33 \pm 0.01$ | - | $0.11 \pm 0.01$ | $40 \pm 4$ |
| 2.25 | $N/m_{\text{eff}}$ | $0.36 \pm 0.01$ | - | $0.16 \pm 0.01$ | $130 \pm 40$ |

Likewise, a transfer of hot charge carriers to the respective other material is favored by the NC geometry, as in the process of the Landau Damping hot charge carriers are generated directly at the surface of the plasmonic domain.[15] This could be the reason why the time constants of the carrier-phonon scattering (e.g. in $N/m_{\text{eff}}$) from the electron and hole LSPR in the heterostructured Au/CuS NCs approach each other more closely. Because both domains are in direct contact with each other, the hot charge carriers could be transferred to the respective other domain. Thus, when adding a CuS shell around a pure Au NCs, the electron phonon scattering is significantly accelerated.

## 3. Conclusion

In this work we synthesized dualplasmonic Au/CuS NCs consisting of Au cores and an asymmetric CuS shell with high yields (83.4 %). In steady-state UV-Vis-NIR spectra two LSPRs are visible at 551 nm (2.25 eV) and 1051 nm (1.18 eV), but give no clear indication of coupling between both LSPRs. Using TAS we show that a resonant excitation at either LSPR maximum in the heterostructured Au/CuS NCs leads to a characteristic hot charge carrier relaxation associated with charge carriers stemming from **both** LSPRs. The almost identical decay dynamics of the hot charge carriers for both excitation energies indicates that 600 fs after exciting the system, it contains the same hot charge carriers (electrons and holes), regardless of the excitation energy. The energy and/or charge transfer between the two Au and CuS domains is therefore very likely based on charge carrier interaction, which is faster than the time range for phononic or thermal interactions. As a result, the time constant of the carrier-phonon scattering of Au NCs decreases (1.6 ps in pure Au NCs and 0.33 ps in Au/CuS hetero NCs). This work presents the first observation of a bidirectional interaction between two LSPRs by transient observation of the hot charge carrier response of both materials. These findings may be used in photocatalysis and other hot charge carrier related applications, especially, because hot charge carriers can be generated in Au NCs, without a resonant excitation of the Au LSPR being needed.

## 4. Experimental

*Used Chemicals*

Chloroform (99.8 %), Copper(II)acetylacetonate (Cu(acac)$_2$, 99.99 %), 1,2-dichlorobenzene (DCB, 99 %), methanol (MeOH, 99.8 %), 1-octadecene (ODE, 90 %), Oleylamin (OLAM, 70 %), Sulfur (S, 99.98 %) and toluene (≥99.5 %) were purchased from Sigma Aldrich. Ethanol (EtOH, 99.8 %) was purchased from Roth. Hydrogen tetrachloroaurate(III) trihydrate (HAuCl$_4$·3 H$_2$O, 99.99 %) was purchased from Alfa Aesar. All chemicals were used as received without further purification.

*Synthesis of Au NCs*

Au NCs were synthesized following a procedure of Sun *et al.*[20] OLAM (20 mL) were heated to 160 °C while stirring under Ar atmosphere and kept at this temperature for 20 min. A 1 M solution of HAuCl$_4$ in water (0.4 mL) was added quickly. The reaction was kept at 160 °C for another 60 min and then cooled to room temperature. After adding MeOH (approx. 20 mL), the reaction solution was centrifuged for 10 min at 10621 rcf. The nanocrystals were dispersed in DCB (20 mL) for the growth of CuS or toluene (20 mL) for the NC mixture.

*Synthesis of heterostructured Au/CuS NCs*

The heterostructured Au/CuS NCs were synthesized following a procedure of Sun *et al*.[20] Cu(acac)$_2$ (0.4 mmol), OLAM (4 mL) and Au NCs in DCB (20 mL) were heated to 80 °C while stirring. A 0.1 M solution of S in DCB (4 mL) was injected quickly. The solution was heated to 100 °C under Ar atmosphere and kept at that temperature for 30 min. After cooling to room temperature EtOH (approx. 20 mL) were added and the reaction solution was centrifuged for 20 min at 3773 rcf and the nanocrystals were dispersed in toluene (20 mL). To separate bigger nanocrystals, the dispersion was centrifuged again at 100 rcf for 5 min and the solution decanted.

*Synthesis of CuS NCs*

The CuS NCs were synthesized following a modified procedure of *Sun et al*.[20] Cu(acac)$_2$ (0.4 mmol), and OLAM (4 mL) were heated to 80 °C while stirring. A 0.1 M solution of S in DCB (4 mL) was injected quickly. The solution was heated to 100 °C under Ar atmosphere and kept at that temperature for 30 min. After cooling to room temperature EtOH (approx. 20 mL) were added and the reaction solution was centrifuged for 20 min at 3773 rcf and the nanocrystals were dispersed in Toluene (20 mL).

*Preparation of the Au and CuS NC mixture*

0.465 mL of the Au NC dispersion were added to 0.4 mL of the CuS NC dispersion to obtain the same absorbance amplitudes as measured for the heterostructured Au/CuS NCs.

*Optical Spectroscopy*

UV-Vis-NIR spectra were collected in transmission mode using a Cary 5000 spectrophotometer by Agilent Technologies. The samples were diluted in toluene and placed in quartz glass cuvettes with 10 mm pathlength. A gratin and detector change is conducted at 800 nm and a lamp change is conducted at 350 nm.

*Transient Absorption Spectroscopy*

Samples were prepared as colloidal solutions in toluene in quartz glass cuvettes with a 2 mm pathway and were constantly stirred in order to prevent laser mediated phase transitions or decomposition.[34] Ultrafast charge carrier dynamics were studied by broadband pump-probe spectroscopy in a set-up described previously and briefly discussed here.[35] 100 fs laser pulses with a wavelength of 800 nm are generated by a Ti:sapphire amplifier system (Spectra-Physics, Spitfire ACE) and split in order to get pump and probe (90:10) pulses. The pump beam wavelength is adjustable from 300 nm to 1600 nm by nonlinear frequency mixing in an optical parametric amplifier and second harmonics generation system (TOPAS, Lightconversion). To ensure comparable TA measurements, all samples were excited at a pump fluence of 150 μJ/cm². At this photon flux, the transient signals behave linearly, so that two-photon processes are ruled out, as shown in **Figure S7**. The probe pulse is converted to a broadband supercontinuum using nonlinear processes in a $CaF_2$ or sapphire crystal, allowing probing from 320 nm to 1600 nm (Ultrafast Systems, Helios FIRE). The probe pulse can be delayed up to 8 ns after photoexcitation *via* an automatic delay line. Pump and probe pulse pass the cuvette with a maximum overlap. After transmission, the probe pulse is measured by a fiber-coupled detector array at different delay times to follow ultrafast processes causing changes in the absorption.

*Electron Microscopy*

TEM images were obtained using a FEI Tecnai G2 F20 device equipped with a field emission gun. The acceleration voltage is 200 kV. The samples were washed by precipitation with EtOH and centrifugation. The nanocrystals were dispersed in chloroform and 20 μl of this dispersion were dropcasted on a 300 mesh Cu grid by Quantifoil.

*Atomic Absorption Spectroscopy*

The gold and copper concentrations of the NC dispersions were determined by atomic absorption spectroscopy (AAS) using a Varian AA 140 spectrometer. Aliquots of the dispersions were taken and the solvent was completely evaporated. Subsequently, the residual solid was dissolved in freshly prepared aqua regia. These solutions were diluted with deionized water. The extinctions in the atomic absorption spectrometer were measured in relation to separately prepared calibration standards with known copper and golds mass concentration.

**Supporting Information**

Supporting Information is available from the Wiley Online Library or from the author.

**Acknowledgements**

P. B. and A. N. contributed equally to this work. The authors would like to thank Armin Feldhoff for providing the XRD facility. P. B. is grateful for being funded by the Hannover School for Nanotechnology (HSN). D.D. thanks the Deutsche Forschungsgemeinschaft for the DFG Research Grant 1580/5-1. D.D. and J.L. are thankful for funding by the Deutsche Forschungsgemeinschaft (DFG, German Research Foundation) under Germany's Excellence Strategy within the Cluster of Excellence PhoenixD (EXC2122, Project ID 390833453). J.L. is grateful for funding through the Caroline Herschel program of Leibniz Universität Hannover. D.K. is grateful for funding by the Konrad-Adenauer-Stiftung (KAS).

Received: ((will be filled in by the editorial staff))

Revised: ((will be filled in by the editorial staff))

Published online: ((will be filled in by the editorial staff))

**Interacting Plasmons:** Heterostructured Au/CuS nanocrystals (NCs) exhibit localized surface plasmon resonance (LSPR) centered at two different wavelengths. By applying ultrafast transient absorption spectroscopy, we observe a characteristic hot charge carrier relaxation associated with both LSPRs regardless of which LSPR is excited beforehand. A transient hot charge carrier response from a non-excited LSPR is not known to literature yet and leads to deeper understanding of LSPRs and their coupling.

*Patrick Bessel[§], André Niebur[§], Daniel Kranz, Jannika Lauth\*, Dirk Dorfs\**

# Probing Bidirectional Plasmon-Plasmon Coupling-Induced Hot Charge Carriers in Dualplasmonic Au/CuS Nanocrystals

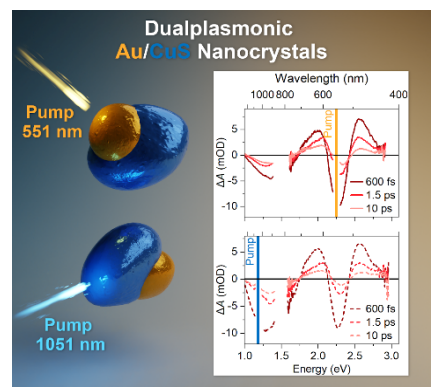

# Supporting Information

**Probing Bidirectional Plasmon-Plasmon Coupling-Induced Hot Charge Carriers in Dualplasmonic Au/CuS Nanocrystals**

*Patrick Bessel, André Niebur, Daniel Kranz, Jannika Lauth\*, Dirk Dorfs\**

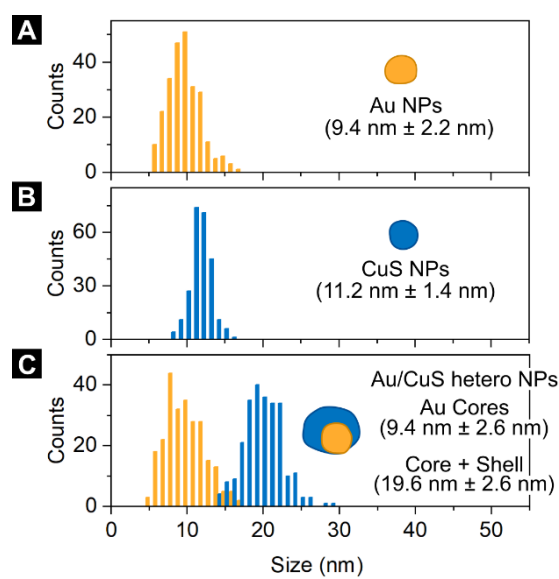

Figure S1: Size distribution determined by measuring 250 NCs on TEM imgaes of (A) single Au NCs (B) single CuS NCs. (C) Diameter of the Au core and diameter of the whole heterostructured Au/CuS NC.

*X-ray diffraction (XRD)*

XRD measurements were performed with a Bruker D8 Advance device in reflection mode, operated at 30 kV and 40 mA using Cu Kα radiation. The samples were washed by precipitation with EtOH and centrifugation. The nanocrystals were dispersed in chloroform and this dispersion were dropcasted onto single crystalline silicon wafer disks.

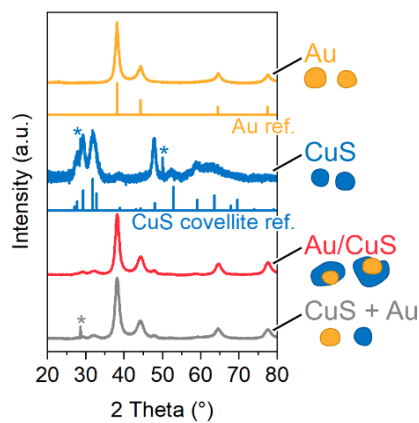

Figure S2: X-ray diffraction patterns of single Au NCs (yellow), single CuS NCs (blue) heterostructured Au/CuS NCs (red) and a mixture of Au NCs and CuS NCs (grey). The obtained crystal phases are elementary gold (PDF Card # 01-071-3755) and covellite (PDF Card # 00-006-0464). Minor impurities, marked with an asterisks, visible by sharp reflections, are found in CuS NC and the mixture of CuS and Au NCs.

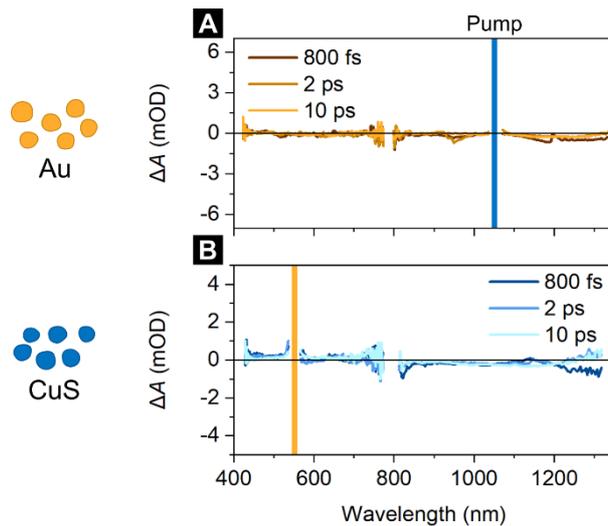

Figure S3: Transient absorption of Au and CuS NCs when excited non-resonantly. Au NCs at 1051 nm (1.18 eV) and CuS NCs at 551 nm (2.25 eV). For both cases no transient response is detected.

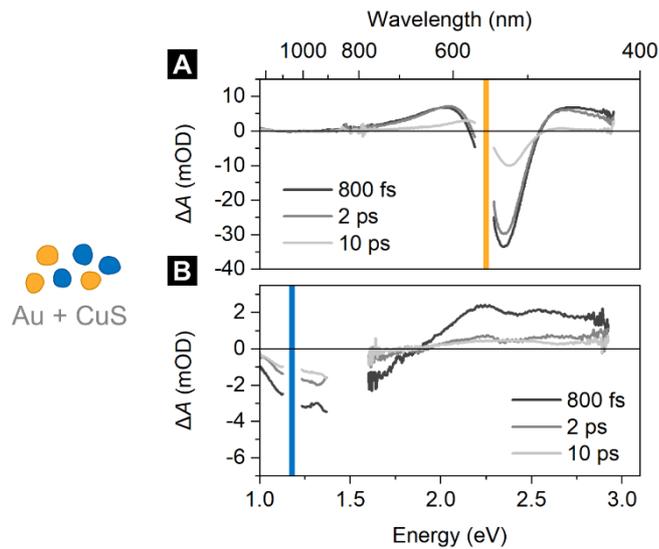

Figure S4: Transient absorption of a mixture of Au NCs and CuS NCs when excited resonantly at (A) 551 nm and (B) CuS NCs at 1051 nm. In both cases, exclusively the transient response of the hot charge carriers stemming from the resonantly excited LSPR are observed. This shows, that the interaction observed in Au/CuS heterocrystals is dependent on a shared interface between Au and CuS.

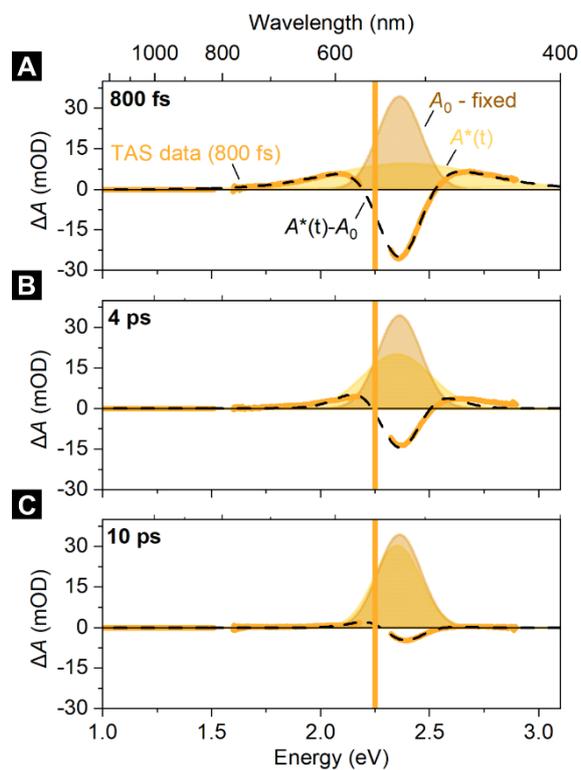

Figure S5: (A) TAS of resonantly excited Au NCs after 800 fs. The dotted line indicates the fitted data as a difference of the excited and a non-excited contribution ($A^*$, yellow and $A_0$, brown, respectively). (B, C) Same data analysis with times of 4 ps and 10 ps after excitation. $A_0$ is kept constant for each time step.

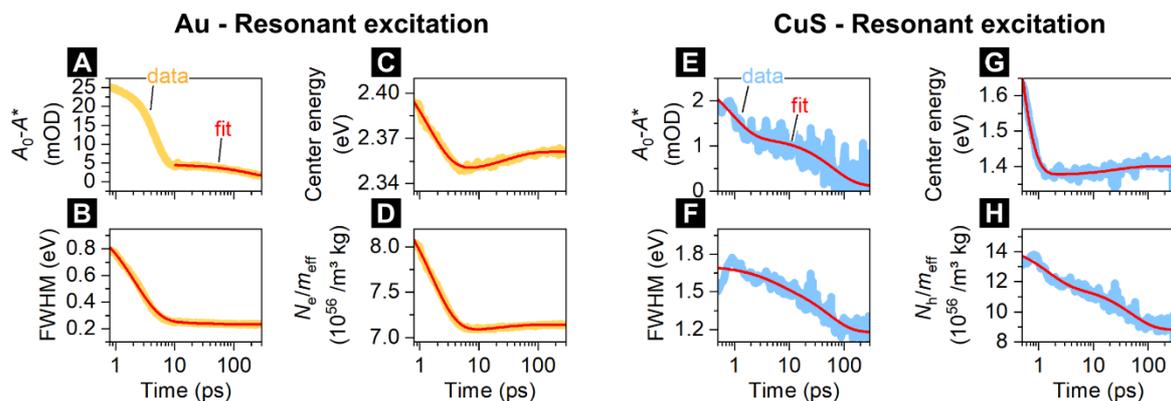

Figure S6: Decay dynamics of the parameters of the Gaussian Curve corresponding to the absorption band after excitation. The dynamics are fitted by the sum of two first order kinetics (exponential decay). The data is derived from TAS of pure Au NCs when excited resonantly at 551 nm and CuS NCs at 1051 nm.

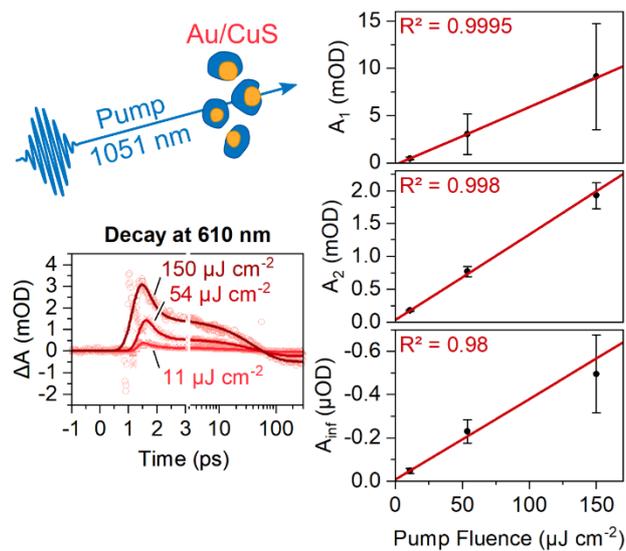

Figure S7: Transient absorption of hetero structured Au/CuS NC when excited resonantly with the LSPR in the CuS domain at 1051 nm (1.18 eV) pulses at different fluences. Fitting of the temporal slice at 610 nm shows, that the amplitude of every process ($A_1$, $A_2$, $A_{inf}$) has a linear dependency.